\documentclass[twocolumn,prl,floatfix,showkeys,showpacs]{revtex4}
\usepackage[colorlinks,linkcolor={blue},citecolor={blue},urlcolor={blue}]{hyperref}

\usepackage[]{graphicx}

\begin{document}

\title{Universal fluctuations in the relaxation of structural glasses}

\author{ Azita Parsaeian and Horacio E. Castillo }
\affiliation{ Department of Physics and Astronomy, Ohio University,
  Athens, OH, 45701, USA } 
\date{\today} 

\begin{abstract}
The presence of strong local fluctuations -- dynamical heterogeneities
-- has been observed near the glass transitions of a wide variety of
materials. Here we explore the possible presence of {\em
  universality\/} in those fluctuations. We compare the statistical
properties of fluctuations obtained from numerical simulations of four
different glass-forming systems: two polymer systems and two particle
systems. We find strong evidence for universality, both in
the qualitative behavior of the fluctuations and in the remarkable
agreement of the scaling functions describing them.
\end{abstract}

%
%
%

\maketitle

{\em Dynamical heterogeneities\/}, i.e. strong fluctuations associated
with nanometer-scale regions of molecules rearranging at very
different rates compared with the bulk~\cite{Ediger_review00,
  Sillescu_review99}, have been observed in a wide variety of
glass-forming systems, from small molecules, to polymers, to network
glasses, to colloidal glasses~\cite{Ediger_review00,
  Sillescu_review99, berthier_bks-silica-dynamics_arXiv07052783,
  Weeks-Weitz}.
A detailed theoretical explanation for those strong fluctuations is
not yet available, although several ideas have been
proposed~\cite{Garrahan-Chandler_geometric-scaling-dynhet_prl-89-035704-2002,
    xia-wolynes_theory-heterogeneity-supercooled_prl-86-5526-2001, 
biroli-bouchaud_upper-critical-dim_epl-67-21-2004,
Toninelli-Wyart-Berthier-Biroli-Bouchaud_pre-71-041505-2005, cccki-prb}.
A growing experimental and numerical literature has been uncovering
various aspects of dynamical heterogentities~\cite{Ediger_review00,
  Sillescu_review99, berthier_bks-silica-dynamics_arXiv07052783,
  Weeks-Weitz, Courtland-Weeks-jphysc03, wang-song-makse,
  castillo-parsaeian_shortrhoC_naturephys-3-26-2007,
  parsaeian-castillo_shortcorr_condmat-0610789,
  parsaeian-castillo_wca_condmat-0802.2560,
  chaudhuri-gao-berthier-kilfoil-kob_random-walk-hetdyn-attracting-colloid_arxiv-07120887,
  Glotzer_etal_JChemPhys, Biroli_etal, dauchot-marty-biroli_prl2005},
but the question of {\em universality\/} --to what degree these
fluctuations behave in the same way in different kinds of glassy
systems-- remains open.

Another common feature of glass-forming materials is {\em physical
  aging\/}~\cite{Hodge95}: for temperatures below the glass
transition, the material falls out of equilibrium, and quantities
probing the system at two times, a ``waiting time'' $t_w$ and a
``final time'' $t$, with $t > t_w$,
depend on {\em both\/} times $t_w$ and $t$, and not just on their
difference $t-t_w$.
In particular, probability distributions and spatial correlations
describing dynamical heterogeneities also show 
aging~\cite{Courtland-Weeks-jphysc03}. Their time
dependences~\cite{wang-song-makse,
  castillo-parsaeian_shortrhoC_naturephys-3-26-2007,
  parsaeian-castillo_shortcorr_condmat-0610789,
  parsaeian-castillo_wca_condmat-0802.2560} display scaling as a
function of the two-time correlation 
$C_{\mbox{\scriptsize
    global}}(t,t_w) \equiv \frac{1}{N}\sum_{j=1}^{N}\exp(i{\bf
  q}.({\bf r_j}(t)-{\bf r_j}(t_w)))$, 
as predicted by a
theoretical framework based on the presence of 
local fluctuations in the age of the sample~\cite{rpg_prl, rpg_prg,
  cccki-prb, ckcc, chamon-charb-cug-reich-sellito_condmat04}.

In this work we examine the question of {\em universality } in the
fluctuation behavior of structural glasses.  We perform numerical
simulations in four different models of glass-forming systems in the
aging regime, and take advantage of the presence of scaling to
quantitatively compare the properties of fluctuations in all of
them. As a result of those comparisons, we find strong evidence that
dynamical heterogeneities do indeed exhibit universal behavior.

We consider a system of polymers (labeled $lj_p$) with Lennard Jones
(LJ) interactions between the monomers, a system of polymers (labeled
$w_p$) with purely repulsive Weeks-Chandler-Andersen (WCA)
interactions~\cite{allen-tildesley, weeks-chandler-andersen}, a system
of particles (labeled $lj_m$) with LJ interactions, and a system of
particles (labeled $w_m$) with WCA interactions.
In our polymer models, nearest-neighbor monomers along a chain are
connected by a FENE anharmonic spring potential, and we use the
Nose-Hoover method to simulate at a constant pressure and
temperature~\cite{allen-tildesley}. Each system, composed of $800$
chains of $10$ monomers each, is equilibrated at a high temperature of
$T_i=5.0$ and then it is instantaneously quenched to a final
temperature $T_f$. All temperatures are measured in units of the
energy scale $\epsilon$ of the LJ or WCA potential. The time of the
quench is taken as the origin of times $t=0$. After the quench the
systems are allowed to evolve for $10^5$ LJ time units. $T_f$ is
chosen low enough that the systems keep aging during the whole low
temperature part of the simulation: $T_f=0.4$ for $w_p$, and $T_f=0.6$
for $lj_p$. Our results are an average over $100$ (resp. $800$)
independent runs for the $lj_p$ (resp. $w_p$) system.
The simulations in the particle systems are as described in
Refs.~\cite{ castillo-parsaeian_shortrhoC_naturephys-3-26-2007,
parsaeian-castillo_wca_condmat-0802.2560}.

\begin{figure}[ht]

  \begin{center}
    \includegraphics[width=3.5in]{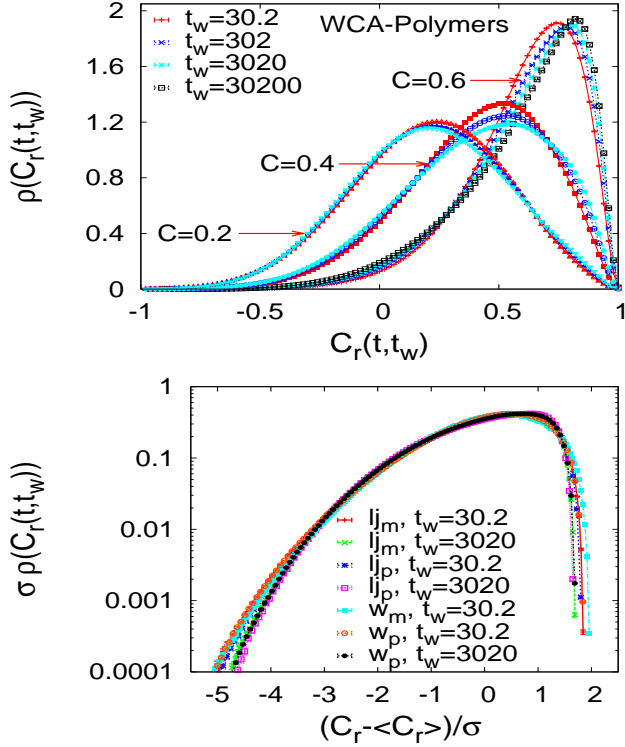}

    \caption{ Probability distribution $\rho(C_{\bf r})$, for coarse
      graining regions containing on average $6.6$ particles. {\em Top
        panel:\/} $\rho(C_{\bf r})$ in the aging regime of a WCA polymer
      glass when $C_{\mbox{\scriptsize global}}(t,t_w)=0.2,0.4,
      0.6$. {\em Bottom panel:\/} Rescaled probability distributions
      $\sigma_C \rho(C_{\bf r})$ as functions of the normalized
      fluctuation $(C_{\bf r}-C_{\mbox{\scriptsize global}})/\sigma_C$
      for $C_{\mbox{\scriptsize global}}(t,t_w)=0.5$, for the four
      systems, with $q = 7.24, 7.37, 7.20, 7.20$ for $w_p, lj_p,
      w_m, lj_m$ respectively. Here $t_w = 30.2, 3020$ in all cases
      except $t_w = 30.2$ only for WCA particles. }

    \label{fig:prob-cdr}

  \end{center}

\end{figure}

\begin{figure}[ht]
  \begin{center} 
    \includegraphics[width=3.5 in]{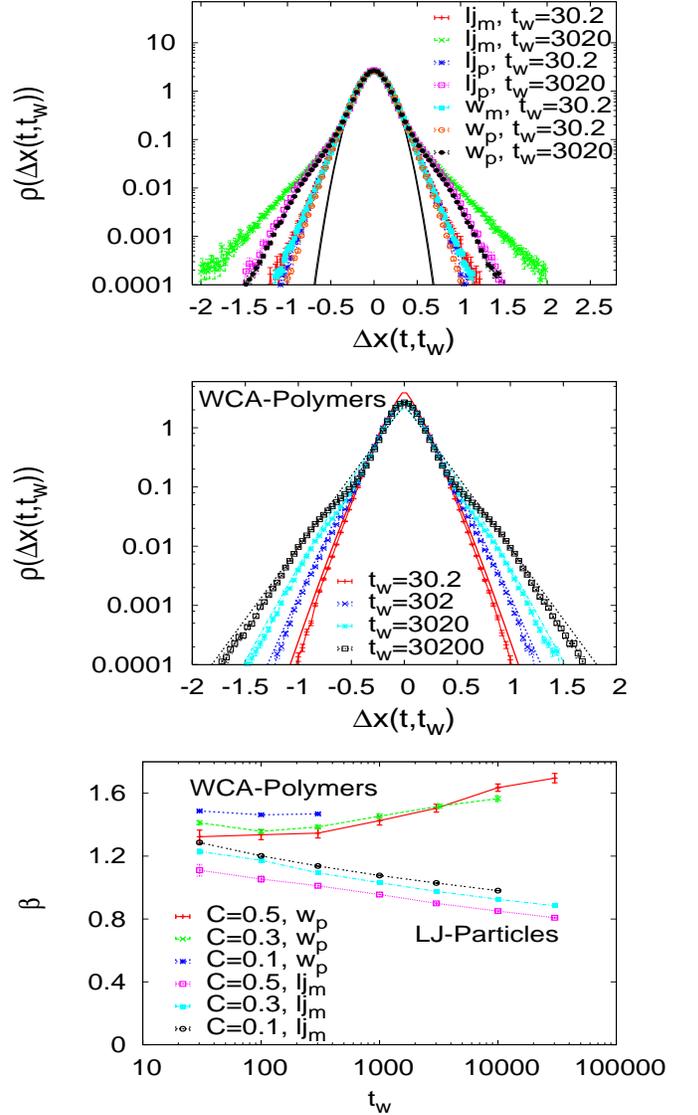}
    \caption{ {\em Top panel:\/} Non-universality of the time evolution
      of the tails of $\rho(\Delta x)$. Plot of $\rho(\Delta x)$ for
      $C_{\mbox{\scriptsize global}}(t,t_w)=0.5$ for the four systems,
      with $t_w= 30.2, 3020$ (except $t_w = 30.2$ only for WCA
      particles). Plotted with a logarithmic vertical axis to
      emphasize the tails of the distributions. Gaussian fits to all
      seven curves are also shown (black full lines); all fits
      collapse with each other. {\em Middle panel:\/} $\rho(\Delta x)$
      with $C_{\mbox{\scriptsize global}}(t,t_w)=0.5$ and $t_w= 30.2,
      302, 3020, 30200$; for a WCA polymer glass. {\em Bottom panel:\/}
      Evolution with waiting time of the $\beta$ exponent describing
      the tails of $\rho(\Delta x)$, for a polymer system and a small
      molecule system, at $C_{\mbox{\scriptsize
          global}}(t,t_w)=0.1,0.3,0.5$. 
    }
    \label{fig:prob-dxi-logs}
  \end{center} 
\end{figure}

\begin{figure}[ht]
\begin{center} 
\includegraphics[width=3.5 in]{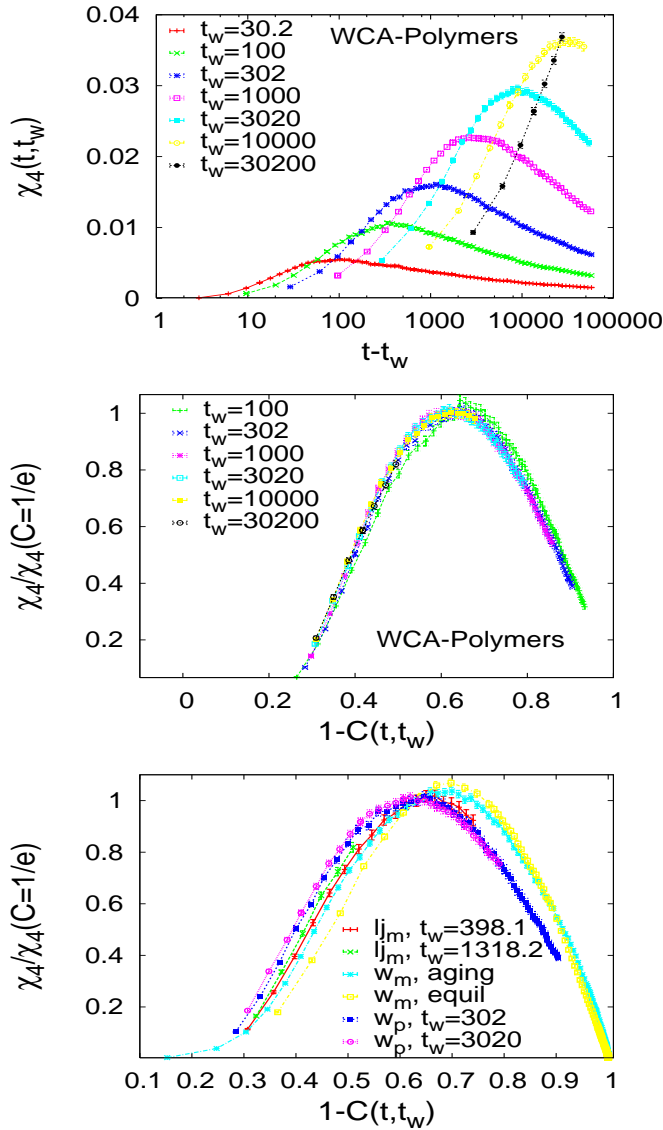}
    \caption{ {\em Top panel:\/} $\chi_4(t,t_w)$ as a function of
      $t-t_w$, with constant $t_w= 100,\cdots,30200$, for WCA polymers
      (data from 5000 independent simulation runs). {\em Middle
        panel:\/} Rescaled $\chi_4$, plotted as a function of $1-C$,
      with constant $t_w= 30.2,\cdots,30200$, for WCA polymers. A very
      good collapse is observed. {\em Bottom panel:\/} Rescaled
      $\chi_4$, plotted as a function of $1-C$, with various waiting
      times, for three of the four systems.}
    \label{fig:chi4}
  \end{center}
\end{figure}

\begin{figure}[ht]
  \begin{center}
    \includegraphics[width=3.5in]{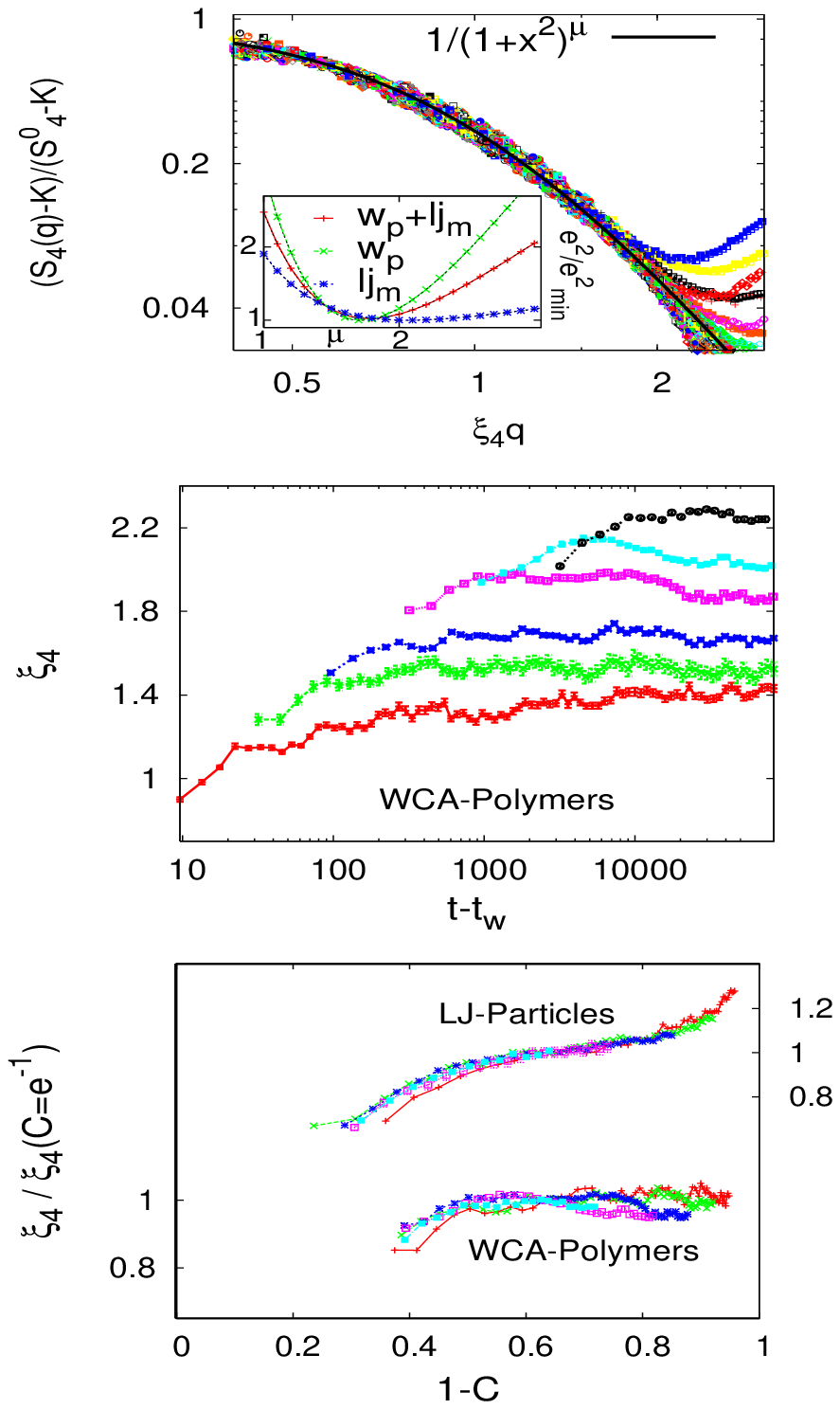}
    \caption{ {\em Top panel:\/} Rescaled $S_4({\bf q}, t, t_w)$, as a
      function of the scaling variable $x= \xi_4(t, t_w) |q|$. Data
      are shown for $165$ time pairs $(t,t_w)$ for LJ particles and
      $338$ time pairs for WCA polymers. {\em Inset:\/} ratios
      of the average squared error $e^2$ divided by its minimum value
      $e^2_{\mbox{\scriptsize min}}$ for $lj_m$ and $w_p$ systems. The curve
      labeled $w_p + lj_m$ shows the average of the two ratios.
      {\em Middle panel:\/} $\xi_4(t, t_w)$ as a function of
      $t-t_w$, with constant $t_w= 100,\cdots,30200$, for WCA
      polymers. The value of $ \xi_4$ at the plateau increases with
      $t_w$. {\em Bottom panel:\/} Rescaled $\xi_4(t, t_w)$, as a
      function of $1-C$. The upper plot is for $LJ$ particles with
      $t_w= 13.2,\cdots,832$ and the lower plot corresponds to $WCA$
      polymers with constant $t_w= 100,\cdots,30200$.}
    \label{fig:xi4}
  \end{center}
\end{figure}

We present results for the probability distributions of 
observables which probe local fluctuations in small regions
of the system: the local coarse grained two-time correlation
function~\cite{castillo-parsaeian_shortrhoC_naturephys-3-26-2007,
  parsaeian-castillo_wca_condmat-0802.2560} $ C_{{\bf r}}(t,t_w)$
and the particle displacements along one
direction $ \Delta x_j(t,t_w) = x_j(t) - x_j(t_w)$~\cite{Weeks-Weitz,castillo-parsaeian_shortrhoC_naturephys-3-26-2007,
  chaudhuri-gao-berthier-kilfoil-kob_random-walk-hetdyn-attracting-colloid_arxiv-07120887}. 
In order to probe the spatial correlations of the fluctuations, we
also consider the generalized dynamic susceptibility $\chi_4
\equiv \int {d^3{\bf r}} \; g_4({\bf r}, t,
t_w)$~\cite{parsaeian-castillo_shortcorr_condmat-0610789,
  Glotzer_etal_JChemPhys, Biroli_etal, dauchot-marty-biroli_prl2005},
where $g_4({\bf r}, 
t, t_w)$ is a $4$-point ($2$-time, $2$-position) correlation
function~\cite{Glotzer_etal_JChemPhys, 
  parsaeian-castillo_shortcorr_condmat-0610789}.

The top panel of Fig.~\ref{fig:prob-cdr} shows the probability
distributions $\rho(C_{\bf r})$ of the local two-time correlation for
$C_{\mbox{\scriptsize global}}(t,t_w) = 0.2, 0.4, 0.6$, 
for WCA polymers. 
As in particle
systems~\cite{castillo-parsaeian_shortrhoC_naturephys-3-26-2007,
  parsaeian-castillo_wca_condmat-0802.2560}, the probability
distributions are approximately independent of $t_w$, for a fixed value
of $C_{\mbox{\scriptsize global}}(t,t_w)$. In the bottom panel of
Fig.~\ref{fig:prob-cdr} we plot the rescaled probability distributions
$\sigma_C \rho(C_{\bf r})$ versus the normalized fluctuation $(C_{\bf
  r}-C_{\mbox{\scriptsize global}})/\sigma_C$ in the one-point
two-time correlator for $C_{\mbox{\scriptsize global}}(t, t_w)=0.5$,
and find that the results are approximately the same for all four
systems: $lj_m$, $w_m$, $lj_p$ and $w_p$.

The top panel of Fig.~\ref{fig:prob-dxi-logs} shows the probability
distributions $\rho(\Delta x)$ of particle displacements for the four
systems, for fixed $C_{\mbox{\scriptsize global}}(t,t_w) = 0.5$ and
$t_w=30.2, 3020$. The distributions look similar to those determined
by confocal microscopy in a colloid with repulsive
interactions~\cite{Weeks-Weitz} but rather different from those found
in an attractive
colloid~\cite{chaudhuri-gao-berthier-kilfoil-kob_random-walk-hetdyn-attracting-colloid_arxiv-07120887}. 
All the distributions collapse
together to a common Gaussian shape for smaller $\Delta x$, but depart
from the Gaussian for larger $\Delta x$. For the shorter waiting
times, $t_w=30.2$, the tails of the distributions in the four systems
are very close to each other, but they significantly separate for
longer $t_w$. In all systems the tails of the distributions become
wider as $t_w$ increases; this is shown in more detail for the $w_p$
system in the middle panel of Fig.~\ref{fig:prob-dxi-logs}. In all four
systems, the tails can be fit in the region $|\Delta x | > 0.5$ by a
nonlinear exponential form $\rho(\Delta x ) \approx N \exp(- {|\Delta
  x/a|}^{\beta} )$. However, the bottom panel of
Fig.~\ref{fig:prob-dxi-logs} shows that for WCA polymers, the exponent
{\em increases} at long $t_w$, while for LJ particles, {\em it
  decreases} at long $t_w$.
According
to~\cite{chaudhuri-gao-berthier-kilfoil-kob_random-walk-hetdyn-attracting-colloid_arxiv-07120887}
the tails of the distribution are representative of particles which
have escaped their cages, whereas the peak
of the distributions are due to the particles which vibrate in 
place. 
This suggests that the difference in the
evolution of the tails could be due to the different diffusive
behavior between small molecules and polymers.

We now consider the spatial correlations of dynamical fluctuations.
The top panel of Fig.~\ref{fig:chi4} shows that $\chi_4(t,t_w)$ has a
peak as a function of $t-t_w$, and that this peak's height and
position
grow with $t_w$.
Similar behaviors are observed in experiments in granular
systems~\cite{dauchot-marty-biroli_prl2005} as the area fraction is
increased, and in numerical simulations of supercooled liquids as the
temperature is reduced~\cite{Glotzer_etal_JChemPhys, Biroli_etal}.
In the middle panel, we plot the ratio $\chi_4/\chi_4(\mbox{C=1/e})$
as a function of $1-C = 1 - C_{\mbox{\scriptsize global}}(t,t_w) $,
for WCA polymers; all the curves collapse into a single master
curve. Both behaviors are identical to those found in LJ and WCA
particles~\cite{parsaeian-castillo_shortcorr_condmat-0610789,
parsaeian-castillo_wca_condmat-0802.2560}. In the bottom panel of
Fig.~\ref{fig:chi4}, the same ratio is plotted for LJ particles, WCA
particles and WCA polymers. The scaling function is similar but
apparantly not identical in the polymer and small molecule cases.

By Fourier transforming the correlation function $g_4({\bf r}, t,
t_w)$, we obtain the 4-point dynamic structure factor $S_4({\bf
  q},t,t_w)$~\cite{Glotzer_etal_JChemPhys, Biroli_etal,
  parsaeian-castillo_shortcorr_condmat-0610789}. We fit its small $q$
behavior ($q < 1.9$) with an empirical scaling form: $ S_4({\bf q})=
\left( S^{0}_{4} - K \right) f(\xi_4 |{\bf q}|) + K$, with $f(x)
\equiv 1/(1+x^2)^{\mu}$, for the $w_p$ and $lj_m$ systems. The dynamic
correlation length $\xi_4(t, t_w)$ and the parameters
$S^{0}_{4}(t,t_w)$ and $K(t,t_w)$ are extracted from each fit.
The unknown scaling function $f(x)$ is the same for all the fits. It
has sometimes been assumed~\cite{Glotzer_etal_JChemPhys} that it has
an Ornstein-Zernicke (OZ) form, corresponding to $\mu=1$. More
recently, however, it has been
shown~\cite{Biroli-Reichman-etal_inhomogeneous-mct-prl2006} that the
function $\chi_{\bf q}$, which is believed to have a similar behavior
to $S_4({\bf q})$, {\em does not\/} have an OZ form, and indeed has an
asymptotic behavior $\chi_{\bf q} \sim |{\bf q}|^{-4}$ for large
${\bf q}$. Indeed, forms other than OZ are better at describing the
data in the supercooled regime of various glassy
models~\cite{Biroli_etal}, and they provide significantly better fits
to the aging regime data presented here. By minimizing the average square error over the
whole set of fits, we determine $\mu \approx 1.8$~\cite{error_value}. 
Although our fits are performed only for moderate to small values of
$x = \xi_4 q$, the asymptotic behavior for large argument, $f(x)
\sim x^{-3.6}$, is not far from the prediction
of~\cite{Biroli-Reichman-etal_inhomogeneous-mct-prl2006}.

The top panel of Fig.~\ref{fig:xi4} shows that all the data sets agree
rather well with the scaling function. The middle panel of
Fig.~\ref{fig:xi4} shows the extracted correlation lengths
vs. $t-t_w$ for different waiting times, for WCA polymers.
We see an initial increase in the correlation length in all the
curves, but when the time difference gets larger, the correlation
length either remains constant or it decreases
slightly~\cite{deciding}. As in LJ particles~\cite{parsaeian-castillo_shortcorr_condmat-0610789}, the plateau value of $\xi_4$ is a growing
function of $t_w$. The bottom panel of Fig.~\ref{fig:xi4} shows
plots of the ratio $\xi_4/\xi_4(C=1/e)$ against $1-C$, for the $w_p$
and $lj_m$ systems. In both cases we find that there is a moderately
good collapse, and it appears that $\xi_4$ goes to a nonzero constant
when $(1-C) \to 1$, (i.e. $(t-t_w)/t_w \to \infty$). However, the
scaling function for monomers grows with $1-C$ for $1-C > 0.5$, while
for the polymers it appears to become approximately constant.

In summary, we have explored the possible presence of universality in
the fluctuations of relaxing structural glasses, by comparing
simulation data from two polymer models and two particle models. As we
pointed out before, some differences are observed:
(a) the scaling plots for $\chi_4/\chi_4(C=1/e)$ and
$\xi_4/\xi_4(C=1/e)$ versus $1-C$ show some small discrepancies
between polymers and particles; and
(b) the behavior of the tails of the distributions of displacements,
$\rho(\Delta x(t,t_w))$, show a qualitatively different evolution with
$t_w$ in the small molecule and polymer cases (which might be due to
their different diffusive behaviors). 
However, there is remarkable similarity in the behavior of
fluctuations in all the systems considered, and we find the evidence
for universality to be very strong. In particular, the following
aspects of the fluctuations appear to be universal:
(i) the fact that the probability distributions $\rho(C_r(t,t_w))$
approximately collapse for different waiting times $t_w$ when $C(t,t_w)$
is held constant;
(ii) the very peculiar shape of the scaling function
$\tilde{\rho}((C_{\bf r} - C_{\mbox{\scriptsize global}})/\sigma_C)
\equiv \sigma_C \rho(C_{\bf r})$;
(iii) the qualitative behavior of the 4-point density susceptibility
$\chi_4(t,t_w)$ and the dynamic correlation length $\xi_4(t,t_w)$ as
functions of $t_w$ and $t-t_w$;
and (iv) the fact that the rescaled quantities
$\chi_4/\chi_4(C=1/e)$ and $\xi_4/\xi_4(C=1/e)$ plotted versus $1-C$
approximately collapse for different waiting times $t_w$.

H.~E.~C. thanks L.~Berthier, J.~P.~Bouchaud, L.~Cugliandolo,
S.~Glotzer, N.~Israeloff, M.~Kennett, M.~Kilfoil, D.~Reichman,
E.~Weeks, and particularly C.~Chamon for suggestions and discussions.
This work was supported in part by DOE under grant DE-FG02-06ER46300,
by NSF under grant PHY99-07949, and by Ohio University. Numerical
simulations were carried out at the Ohio Supercomputing Center.
H.~E.~C. acknowledges the hospitality of the Aspen Center for Physics,
were part of this work was performed.


\begin{thebibliography}{99}
\bibitem{error_value}
  The minization yields a value of $\mu = 1.769 \pm 0.001$. The
  statistical error is small, but the systematic errors due to the
  choice of interval and possible differences between monomer and
  polymer systems are more difficult to estimate, and probably much
  larger.
\bibitem{deciding} 
  Deciding between these two possibilities would require even longer time
  spans and larger statistics.
\end{thebibliography}
\end{document}